\documentclass[
  english,
  aps,
  prl,
  reprint,
  floatfix,
  footinbib,
  preprintnumbers,
  longbibliography
]{revtex4-2}

\usepackage{mathtools}
\usepackage{amsmath, amsfonts, amsthm, amssymb, graphicx, color, hyperref, slashed, braket}
\usepackage{pstool}
\usepackage{float}
\usepackage[utf8]{inputenc}
\usepackage[normalem]{ulem}
\usepackage[english]{babel}
\usepackage{blkarray}
\usepackage{mathrsfs}
\usepackage{diagbox}
\usepackage{multirow}
\usepackage{tikz}
\usetikzlibrary{arrows.meta,positioning}
\usepackage{physics}

\usepackage{amsmath,amssymb,color}
\usepackage{diagbox}

\newcommand{\bea}{\begin{eqnarray}}
\newcommand{\eea}{\end{eqnarray}}
\newcommand{\be}{\begin{equation}}
\newcommand{\ee}{\end{equation}}

\newcommand{\dsl}{\pa \kern-0.5em /}

\newcommand{\pa}{\partial}

\newcommand{\nn}{\nonumber\\}

\begin{document}
\preprint {USTC-ICTS/PCFT-26-52}


\title{The explicit Lorentz invariant QED pair production rates \\
from superstrings}


\author{J. X. Lu}
\affiliation{\em  Interdisciplinary Center for Theoretical Study\\
 University of Science and Technology of China, Hefei, Anhui
 230026, China\\
 and\\
 Peng Huanwu Center for Fundamental Theory, Hefei, Anhui 230026, China\\}


\date{\today}

\begin{abstract}
We present a new avenue to the usual Schwinger effect via the open string pair production  for a system of two Dp branes with $1 \le p \le 6$ in Type II superstrings.  The two Dp are  placed parallel at a separation with one of them carrying the most general constant electromagnetic background  allowed for the pair production. By taking the so-called field theory limit, we obtain systematically, from the open string pair production rate, the respective \textit{explicit Lorentz invariant} QED pair production rate in diverse dimensions for a pair of charged/anti-charged massive particles, which can be scalars, spinors or vectors.  

\end{abstract}


\maketitle

\noindent
{\textbf{ Introduction:}} These is so far no clear experimental or observational evidence for quantum gravity (QG).  Nevertheless,  a quantum theory of gravity is needed to understand fundamental questions such as the underlying physics of the cosmological singularity and black hole singularity, and the nature of dark energy.  For this and in the absence of experimental input, various theoretical approaches have been put forward.  Among these, though still facing much debate, string theory stands out as a favorable one for QG.  This has further been reinforced recently by the bottom-up scattering amplitudes bootstrap, utilizing basic principles like Lorentz invariance, locality, and unitarity plus some modest  requirement on the amplitude high energy behavior.  For example, string theory has recently been argued to emerge inevitably or uniquely from a few simple aforementioned assumptions  about physical scattering \cite{Cheung:2024uhn, Cheung:2025tbr}.\\
\indent
       In this Letter, we address a question, from the top-down approach of string theory, on how to obtain the usual quantum field theory results from the computations of superstring theory.  In particular, we focus on how to obtain the usual Schwinger pair production \cite{Schwinger:1951nm} for a pair of charged/anti-charged massive particles which can be scalars, spinors or vectors in diverse dimensions, in the most general constant electromagnetic background  allowed for this purpose and in an explicit Lorentz invariant form, from the open string one for a system of two Dp branes, called one visible and the other hidden,  in Type II superstrings.  The visible Dp carries the same constant electromagnetic flux and is placed parallel at a separation $y$ from the hidden Dp.  In obtaining the respective usual rates of quantum electrodynamics (QED), we need to use  the known field content knowledge of the massive vector multiplet describing the lowest modes of  the open string connecting the two Dp branes from the worldvolume perspective, and take the so-called field theory limit.\\
\indent 
This work is motivated by the author's previous one  for a system of two D3 branes in the same setting except for a somewhat specific background of  collinear constant electric and magnetic fields on the visible D3 \cite{Lu:2023sag}.  As explained there, the open string pair production cannot occur for a single isolated D3 (or in general a single isolated Dp)  if the applied electric field on the brane  is less than the so-called critical one,  due to the virtual open string being unbreakable. The two-D3 system is actually the simplest one for this purpose. The open string pair production comes from the virtual open string and the virtual anti open string connecting the two D3, therefore along the extra dimensions with respect to either D3.   When the field theory limit is taken, the open string pair production rate becomes \cite{Lu:2023sag}
\be\label{eg3pprate-new}
{\cal W}^{({\rm String})} =  \frac{2 (e^{2} E B)}{(2 \pi)^2 } \frac{\left[\cosh\frac{\pi B}{ E} +1\right]^2}{\sinh \frac{\pi B}{E}}  e^{- \frac{  \pi m^{2}}{e E }},
\ee
which counts the contribution from the 16 pairs consisting of the lowest charged/anti charged massive modes of the open string/anti open string pair, with respect to, say, the unbroken U(1) of the visible D3.  Note that each such mode has the same mass  $m = T_{F} y $ with $T_{F} = 1/(2\pi\alpha') $  the fundamental string tension. One can see this by sending the magnetic field $B \to 0$ and the rate (\ref{eg3pprate-new}) becomes
\be
{\cal W}^{({\rm String})}  (B \to 0) =  16\, \frac{(e E)^{2}}{(2 \pi)^{3}}\, e^{- \frac{\pi m^{2}}{e E}},
\ee
where the pre-factor 16 counts just the number of pairs or the number of the lowest massive modes of the open string, noting that the QED rate for a massive scalar pair  is simply
\be 
{\cal W}^{\rm (QED)}_{\rm scalar} =  \frac{(e E)^{2}}{(2 \pi)^{3}}\, e^{- \frac{\pi m^{2}}{e E}}.
\ee
 From the D3 worldvolume perspective, these 16 pairs give the corresponding 5 scalar pairs, 4 spinor pairs and one vector pair.   We therefore expect the following identity in general
\be\label{rate-identity}
{\cal W}^{({\rm String})} = 5\, {\cal W}^{\rm (QED)}_{\rm scalar} + 4\, {\cal W}^{\rm (QED)}_{\rm spinor} + {\cal W}^{\rm (QED)}_{\rm vec.}, 
\ee
where the non-perturbative stringy rate ${\cal W}^{({\rm String})}$ is given in (\ref{eg3pprate-new}) while the known non-perturbative QED rates under the same collinear electric and magnetic fields were computed long time ago by Nikishov  \cite{nikishov} 
 \bea\label{scalar/spinor}
 {\cal W}^{\rm (QED)}_{\rm scalar} &=& \frac{(e E) (e B)}{2 (2\pi)^{2}} {\rm csch} \left(\frac{\pi B}{E}\right) \, e^{- \frac{\pi m^{2}}{e E}},\nn
{\cal W}^{\rm (QED)}_{\rm spinor} &=& \frac{(e E) (e B)}{(2\pi)^{2}} \coth\left(\frac{\pi B}{E}\right) \, e^{- \frac{\pi m^{2}}{e E}},
\eea 
for a pair of scalars and a pair of spinors, and by Kruglov \cite{Kruglov:2001cx}
 \be\label{w-bosonr}
{\cal W}^{\rm (QED)}_{\rm vec.} = \frac{(e E)(e B)}{ 2 (2 \pi)^{2}} \frac{ 2 \cosh \frac{2 \pi B}{E} + 1}{ \sinh \frac{\pi B}{E}} \, e^{- \frac{\pi m^{2}}{e E}},
\ee
for a pair of vectors.  One can check easily that the identity (\ref{rate-identity}) holds indeed if all the masses involved are set equal.

\noindent
\textbf{Generality:} The above hints on how to obtain the explicit Lorentz invariant QED rates for a pair of such particles in $(1 + p)$-dimensions if the corresponding open string rate  is known. One additional piece of information needed  is the known field content of the massive vector multiplet describing the lowest modes of the open string connecting the two D$p$, from the worldvolume perspective. From the bulk spacetime view, the number of these modes is  always 16, consisting of $8_{\rm B}$ bosons and $8_{\rm F}$ fermions.  

Let us pause briefly to understand the 16 pairs contributing to the rate.  Following \cite{note3}, we have  the usual
massless $4 (8_{\rm F} + 8_{\rm B})$ degrees of freedom(DOF) when $y = 0$ for the system of two Dp. Among these,  $2 (8_{\rm F} + 8_{\rm B})$  become massive with their mass $m = T_{\rm F} y$ due to the unbroken SUSY, when $y\neq 0$. This  reflects  $U (2)$  $\to U(1) \times U(1)$ from $y = 0 \to y \neq 0$.  The two broken generators  give  such 16 pairs with each charged and anti charged under, say, the visible brane's unbroken U(1).  From the worldvolume perspective, these 16 pairs count  $(8 - p)$ scalar  pairs,  $2^{3 - [\frac{p + (-1)^{[(p + 2)/4]}}{2}]}$ spinor pairs and one vector pair for $3 \le  p \le 6$.   Here $[a]$ denotes the integral part of $a$. For $1 \le p \le 2$, we  have 8 scalar  and 8 spinor pairs.  The detail worldvolume field content for $1 \le p \le 6$ is given in Table \ref{table1}. The on-shell DOF counting for a massive particle which can be a scalar, a spinor or a vector in d dimensions is given, respectively, as
\be\label{dof-c}
d_{\rm scalar} = 1, \quad d_{\rm spinor} = 2^{[\frac{d}{2}] - 1}, \quad d_{\rm vec.} = d - 1.
\ee

\begingroup
\squeezetable
\begin{table}[!htbp]
\caption{\label{table1} The worldvolume field content of the lowest modes of  the open string connecting the two D$p$ branes}
\begin{center}
\begin{tabular}{|c|c|c|c|}
\hline
  \diagbox{$d = 1 + p$}{Fields} & Scalars& Spinors &Vectors \\
\hline
 7 & 2 & 2 & 1 \\
 \hline
6 & 3 & 2 & 1 \\
\hline
 5 & 4 & 4 & 1 \\
 \hline
 4 & 5 & 4 & 1\\
 \hline
 3 & 8 & 8 & 0\\
 \hline
 2 & 8 & 8 & 0\\
 \hline  
\end{tabular}
\end{center}
\end{table}
\endgroup

The open string pair production  is related to the imaginary part of  the open string one-loop annulus amplitude between the two D$p$, and can be obtained from the residues of an infinite number of simple poles of the integrand of the amplitude in an integral form, following \cite{Bachas:1992bh, Porrati:1993qd, nikishov, Cohen:2008wz, Lu:2023jxe}.  This has been discussed in a great detail in a lengthy paper \cite{Jia:2019hbr} by the present author and his collaborators. A sketch of this is given in \textbf{Appendix}.  
 
 By taking the field theory limit (see \textbf{Appendix} ),  we have the rate from (\ref{pp-rate}) as
  \bea\label{text-field-pp-rate} 
{\cal W}^{(\rm String)}_{p, \,p} &=& \frac{ {\bar\nu}^{\frac{p - 3}{2}}_{0} \, \nu_{1}\, \nu_{2} } { (2 \pi)^{p - 2}}   \frac{\left[\cosh \frac{\pi {\nu}_{1}}{\bar \nu_{0}} + \cosh \frac{\pi \nu_{2}}{\bar \nu_{0}} \right]^{2}}{\sinh\frac{ \pi \nu_{1}}{\bar \nu_{0}} \sinh\frac{\pi\nu_{2}}{\bar \nu_{0}}} \nn
&\,&\times e^{- \frac{ \pi m^{2}}{\bar \nu_{0}}},
 \eea
 where as before $m = y/(2\pi \alpha')$ and
 \be\label{parameter-relation}
 \bar\nu_{0} = e \sqrt{\alpha}, \,\, \nu_{1} = e \sqrt{- \beta}, \,\, \nu_{2} = \sqrt{ - \gamma}
 \ee
 with $\alpha > 0$ and $\beta, \gamma \leq 0$, satisfying  (with $F$ the usual flux on the visible brane)
 \bea\label{abc}
 &&\alpha\beta\gamma = \frac{1}{3!}[{\rm tr} F^{6} - \frac{3}{4} {\rm tr} F^{4}  {\rm tr} F^{2} + \frac{1}{8} ({\rm tr} F^{2})^{3}] \ge 0,\nn
 &&\alpha\beta + \alpha\gamma + \beta\gamma =   - \frac{1}{4} [ {\rm tr} F^{4} - \frac{1}{2} ( {\rm tr} F^{2})^{2}],\nn
 &&\alpha + \beta + \gamma = \frac{1}{2} {\rm tr} F^{2}.
  \eea
 Note also the rate relation between the D$p$ case and the D$(p + 2)$ case, following \cite{Jia:2019hbr} and the discussion given right after (\ref{hat-parameters}), is
 \be\label{text-rate-relation}
 {\cal W}^{\rm (String)}_{p, p} = \lim_{\nu_{a} \to 0} \frac{(2 \pi)^{2}}{\bar\nu_{0}} {\cal W}^{\rm (String)}_{p + 2,\, p + 2}\, ,
 \ee
 where $a = 2$ for $p = 3$  or  $4$, and $a = 1$ for $p = 1$ or $2$.  We will use this relation for consistent checks later on. 
 
 The formula (\ref{text-field-pp-rate}) is our starting point for having the QED rate for a pair of scalars, or  spinors or  vectors in diverse dimensions and in an explicit Lorentz invariant form.  We discuss case by case in what follows.
 
 \noindent
 \textbf{The} $\mathbf {p = 1}$ \textbf{or 2 case:}  For either case, the explicit Lorentz invariant stringy rate can be obtained from (\ref{text-field-pp-rate}) by taking $\nu_{2} \to 0$ (or $\gamma \to 0$)  and $\nu_{1} \to 0$ (or $\beta \to 0$) as
\be
\label{1-2-rate}
{\cal W}^{(\rm String)}_{p, p} = 16 \frac{ \left(e {\sqrt{\alpha}}\right)^{\frac{p + 1}{2}}} { (2 \pi)^{p}} \,e^{- \frac{ \pi m^{2}}{e \sqrt{\alpha}}},
 \ee
 where we have used $\bar\nu_{0} = e \sqrt{\alpha}$ from (\ref{parameter-relation}). The limits $\beta, \gamma \to 0$ are consistent with (\ref{abc}), noting now ${\rm tr} F^{4} = ({\rm tr} F^{2})^{2}/2$ and ${\rm tr} F^{6} = ({\rm tr} F^{2})^{3}/4$.  We have also $\alpha = {\rm tr} F^{2}/2 > 0$.  For either case, we have the worldvolume field content, from Table \ref{table1},  of 8 charged scalars and 8 charged spinors with equal mass and so we have
 \be
 {\cal W}^{(\rm String)}_{p, p} = 8 {\cal W}^{\rm (QED)}_{\rm scalar}  + 8 {\cal W}^{\rm (QED)}_{\rm spinor},
 \ee
 which gives
 \be
 {\cal W}^{\rm (QED)}_{\rm scalar} = {\cal  W}^{\rm (QED)}_{\rm spinor} =  \frac{\left(e {\sqrt{\alpha}}\right)^{\frac{p + 1}{2}}} { (2 \pi)^{p}} \,e^{- \frac{ \pi m^{2}}{e \sqrt{\alpha}}},
 \ee 
 since a scalar or a spinor in either case counts only one DOF.  So in general for $p = 1$, we have the explicit Lorentz invariant rates, respectively, as
 \be
 {\cal W}^{\rm (QED)}_{\rm scalar}  = \frac{e E } {2 \pi} \,e^{- \frac{ \pi m^{2}_{\rm scalar}}{e E}},\, {\cal W}^{\rm (QED)}_{\rm spinor}  = \frac{e E } {2 \pi} \,e^{- \frac{ \pi m^{2}_{\rm spinor}}{e E}}, 
 \ee
 where $m_{\rm scalar}$ and $m_{\rm spinor}$ are the respective masses,   and $\alpha = {\rm tr} F^{2}/ 2 = E^{2}$ with $E$ the Lorentz invariant electric field $E = |F_{01}|$.  They agree  with the respective known rates  \cite{Gavrilov:1996pz, Cohen:2008wz, Huet:2010nt} with the same constant electric field.  For $p = 2$, by the same token, we have now
 \bea\label{3d-rate}
 {\cal W}^{\rm (QED)}_{\rm scalar} &=&  \frac{\left(e {\sqrt{\alpha}}\right)^{\frac{3}{2}}} { (2 \pi)^{2}} \,e^{- \frac{ \pi m_{\rm scalar}^{2}}{e \sqrt{\alpha}}}, \nn
  {\cal W}^{\rm (QED)}_{\rm spinor} &=&  \frac{\left(e {\sqrt{\alpha}}\right)^{\frac{3}{2}}} { (2 \pi)^{2}} \,e^{- \frac{ \pi m_{\rm spinor}^{2}}{e \sqrt{\alpha}}},
 \eea
 where the Lorentz invariant $\alpha = F^{2}/2 = E_{1}^{2} + E^{2}_{2} - (F_{12})^{2} > 0$ with $E_{a} = F_{0a}$. In a particular frame for which $F_{12} = 0$, these rates agree with those given in \cite{Gavrilov:1996pz}. While for a pure constant electric field, the spinor rate agrees with that given in  \cite{Gusynin:1998bt, Allor:2007ei}.  However, to our best of knowledge,  the manifest Lorentz invariant rates (\ref{3d-rate}) have not been given before.  Note that a massive charged vector in 3 dimensions has 2 DOFs, so we have
 \be 
  {\cal W}^{\rm (QED)}_{\rm vec.} = 2 \frac{\left(e {\sqrt{\alpha}}\right)^{\frac{3}{2}}} { (2 \pi)^{2}} \,e^{- \frac{ \pi m_{\rm vec.}^{2}}{e \sqrt{\alpha}}}.
  \ee
  
  \noindent
 \textbf{The} $\mathbf {p = 3}$ \textbf{or 4 case:}  For either case, we have rate from (\ref{text-field-pp-rate}) by taking $\nu_{2} \to 0$ (or $\gamma \to 0$) as
  \bea
 \label{3-4-rate}
{\cal W}^{(\rm String)}_{p, p} &=& \frac{ 2  \left(e \sqrt{\alpha}\right)^{\frac{p - 1}{2}} e \sqrt{|\beta|}} {(2 \pi)^{p - 1}}\, \frac{\left[1 + \cosh \pi\sqrt{\frac{|\beta|}{\alpha}} \right]^{2}}{\sinh\pi \sqrt{\frac{|\beta|}{\alpha}}} \nn
&\,&\times e^{- \frac{ \pi m^{2}}{e\sqrt{\alpha}}}.
 \eea   
Now one can show  ${\rm tr} F^{6} = \frac{3}{4} {\rm tr} F^{4}  {\rm tr} F^{2} + \frac{1}{8} ({\rm tr} F^{2})^{3}$, being consistent with $\gamma = 0$. We can then solve the remaining equations (\ref{abc}) to give explicitly the Lorentz invariants
  \bea\label{alpha-beta-r}
  \alpha &=& \frac{1}{2}\left[\frac{1}{2} {\rm tr} F^{2} + \sqrt{{\rm tr} F^{4} - \frac{1}{4} ({\rm tr} F^{2})^{2}}\right], \nn
   \beta &=& \frac{1}{2}\left[\frac{1}{2} {\rm tr} F^{2} - \sqrt{{\rm tr} F^{4} - \frac{1}{4} ({\rm tr} F^{2})^{2}}\right] .
  \eea
  Note that from (\ref{abc}) and $\gamma = 0$,  $\alpha > 0$ and $\beta \le 0$ imply ${\rm tr} F^{4} - ({\rm tr} F^{2})^{2}/2 \ge 0$.  For $p = 3$ case, we have ${\rm tr} F^{2} = 2 (\vec E^{2} - \vec B^{2})$ and ${\rm tr} F^{4} - ({\rm tr} F^{2})^{2}/4   = (\vec E^{2} - \vec B^{2})^{2} + 4 (\vec E \cdot \vec B)^{2}$ where $\vec E$ and $\vec B$ are the usual electric and magnetic vector fields.  So we have
 \bea
  &&\alpha = \frac{\vec E^{2} - \vec B^{2}}{2} + \frac{1}{2} \sqrt{\left(\vec{E}^{2} - \vec{B}^{2}\right)^{2} + 4(\vec E \cdot \vec B)^{2}}, \nn
   &&\beta = \frac{\vec E^{2} - \vec B^{2}}{2} - \frac{1}{2}\sqrt{\left(\vec E^{2} - \vec B^{2}\right)^{2}  + 4 (\vec E \cdot \vec B)^{2}},\qquad
  \eea 
and from (\ref{3-4-rate})  the rate 
   \bea
 \label{3-rate}
&&{\cal W}^{(\rm String)}_{3, 3} = \frac{e^{2} \sqrt{\alpha |\beta|}} {(2 \pi)^{2}}\, e^{- \frac{ \pi m^{2}}{e\sqrt{\alpha}}}\nn
&&\times \frac{\frac{5}{2} + 4 \cosh \pi\sqrt{\frac{|\beta|}{\alpha}} + \cosh 2 \pi\sqrt{\frac{|\beta|}{\alpha}} + \frac{1}{2}}{\sinh\pi \sqrt{\frac{|\beta|}{\alpha}}}. 
 \eea    
  It is known that in the field theory limit,  we have 5 scalar pairs, 4 spinor pairs and one vector pair, therefore expect to have the identity (\ref{rate-identity}).
  By setting $\vec B = 0$ ($\beta = 0$), we have, from (\ref{3-rate}) and (\ref{dof-c}), the following expected ones, 
    \bea\label{3-rates-special}
  {\cal W}^{\rm (String)}_{3, 3} &=& 16 \,\frac{(e E)^{2}}{(2 \pi)^{3}} \, e^{- \frac{ \pi m^{2}}{e E}} = 16\, {\cal W}^{\rm (QED)}_{\rm scalar},\nn
   {\rm W}^{\rm (QED)}_{\rm vec.} &=&  3   {\cal W}^{\rm (QED)}_{\rm scalar}, \, \, {\cal W}^{\rm(QED)}_{\rm spinor} = 2 {\cal W}^{\rm (QED)}_{\rm scalar}.\quad
    \eea
  Combining (\ref{3-rate}), (\ref{rate-identity}) and (\ref{3-rates-special}), we have the Lorentz invariant rates for a pair of scalars,  a pair of spinors and a pair of vectors, respectively,  as
  \bea\label{3-rates}
  &&{\rm W}^{\rm (QED)}_{\rm vec.} = \frac{e^{2} \sqrt{\alpha |\beta|}} { 2 (2 \pi)^{2}} \frac{ 1  + 2 \cosh 2 \pi\sqrt{\frac{|\beta|}{\alpha}}}{\sinh\pi \sqrt{\frac{|\beta|}{\alpha}}} \,e^{- \frac{ \pi m_{\rm vec.}^{2}}{e\sqrt{\alpha}}},\nn
   && {\cal W}^{\rm (QED)}_{\rm scalar}  =  \frac{e^{2} \sqrt{\alpha |\beta|}} { 2 (2 \pi)^{2}}\, {\rm csch}\pi\sqrt{\frac{|\beta|}{\alpha}} \,e^{- \frac{ \pi m_{\rm scalar}^{2}}{e\sqrt{\alpha}}}, \nn
 && {\cal W}^{\rm(QED)}_{\rm spinor} = \frac{e^{2} \sqrt{\alpha |\beta|}} {(2 \pi)^{2}} \coth  \pi\sqrt{\frac{|\beta|}{\alpha}}\,e^{- \frac{ \pi m_{\rm spinor}^{2}}{e\sqrt{\alpha}}}.\quad
 \eea 
 One can check that the above rates agree with the respective rates (\ref{scalar/spinor}) for a scalar pair and a spinor pair  \cite{nikishov}  (see also \cite{Daugherty:1976mg, Gavrilov:1996pz, Kim:2003qp, Korwar:2018euc, Dunne:2025cyo}) and  (\ref{w-bosonr})  for a vector pair \cite{Kruglov:2001cx}  when the electric field and the magnetic field are collinear, i.e., $\vec E \cdot \vec B = \pm E B$, a non-trivial test!  Further, the explicit Lorentz invariant ${\cal W}^{\rm (QED)}_{\rm scalar}$ and ${\cal W}^{\rm (QED)}_{\rm spinor}$ (\ref{3-rates}) are also known \cite{Korwar:2018euc, Cho:2000ei, Hattori:2023egw, Gupta:2025srv} but the Lorentz invariant ${\cal W}^{\rm (QED)}_{\rm vec.}$ has not been given before and is new, to the best of our knowledge. 
 
 For $p = 4$ case,  we have from (\ref{3-4-rate}) the following
  \bea
 \label{4-rate}
&&{\cal W}^{(\rm String)}_{4, 4} = \frac{\left(e \sqrt{\alpha}\right)^{\frac{3}{2}} e \sqrt{|\beta|}} {(2 \pi)^{3}} \,e^{- \frac{ \pi m^{2}}{e\sqrt{\alpha}}} \nn
&&\times \frac{2 +  4 \cosh \pi\sqrt{\frac{|\beta|}{\alpha}} + \cosh 2\pi\sqrt{\frac{|\beta|}{\alpha}} + 1}{\sinh\pi \sqrt{\frac{|\beta|}{\alpha}}} ,\quad
 \eea   
 which is expected, from Table \ref{table1}, to satisfy the following relation
 \be\label{4-string-QED}
 {\cal W}^{\rm (String)}_{4, 4} = 4 {\cal W}^{\rm (QED)}_{\rm scalar}  + 4 {\cal W}^{\rm(QED)}_{\rm spinor} + {\rm W}^{\rm (QED)}_{\rm vector}.
 \ee
 By setting $F_{ab} = 0$ ($\beta = 0$) with $a, b = 1, \cdots 4$, we have from (\ref{4-rate}) and (\ref{dof-c}) 
 \bea\label{4-QED-relation}
 {\cal W}^{\rm (String)}_{4, 4} &=& 16  \frac{(e E)^{5/2}}{(2\pi)^{4}} \, e^{- \frac{\pi m^{2}}{e E}} = 16 \, {\cal W}^{\rm (QED)}_{\rm scalar},\nn
  {\cal W}^{\rm (QED)}_{\rm spinor} &=& 2 {\cal W}^{\rm (QED)}_{\rm scalar}, \, {\cal W}^{\rm (QED)}_{\rm vec.} = 4 {\cal W}^{\rm (QED)}_{\rm scalar}, 
 \eea 
 where  $E = |\vec E|$ with $E_{a} = F_{0 a}$. See \cite{note1} for comparing the rates (\ref{4-QED-relation}) with the known results. 
  
 Combing (\ref{4-rate}), (\ref{4-string-QED}) and (\ref{4-QED-relation}), we can read the Lorentz invariant rates for a pair of scalars, a pair of spinors and a pair of vectors, respectively, as
 \bea \label{4-rates}
&& {\cal W}^{\rm (QED)}_{\rm vec.}= \frac{\left[e \sqrt{\alpha}\right]^{\frac{3}{2}} e {|\beta|}^{\frac{1}{2}}} {(2 \pi)^{3}}\frac{\cosh 2\pi \sqrt{\frac{|\beta|}{\alpha}}+ 1}{\sinh\pi \sqrt{\frac{|\beta|}{\alpha}}} e^{- \frac{ \pi m_{\rm vec.}^{2}}{e\sqrt{\alpha}}},\nn
&&{\cal W}^{\rm (QED)}_{\rm spinor} = \frac{ \left[e \sqrt{\alpha}\right]^{\frac{3}{2}} e {|\beta|}^{\frac{1}{2}}}{(2 \pi)^{3}} \coth \pi\sqrt{\frac{|\beta|}{\alpha}} \, e^{- \frac{ \pi m_{\rm spinor}^{2}}{e\sqrt{\alpha}}},\nn
&&{\cal W}^{\rm (QED)}_{\rm scalar} = \frac{\left[e \sqrt{\alpha}\right]^{\frac{3}{2}} e {|\beta|}^{\frac{1}{2}}} { 2 (2 \pi)^{3}} {\rm csch} \pi \sqrt{\frac{|\beta|}{\alpha}} e^{- \frac{ \pi m_{\rm scalar}^{2}}{e\sqrt{\alpha}}}.\quad
\eea 
Following the prescription given in \cite{note2}, we take the vector-rate relation between the $p = 4$ and the $p = 2$ as an illustration.  We expect to have
\be
\lim_{|\beta| \to 0} \frac{(2\pi)^{2}}{e \sqrt{\alpha}} {\cal W}^{\rm (QED)}_{\rm vec.} = {\cal W}^{\rm (QED)}_{\rm vec.} + 2 {\cal W}^{\rm (QED)}_{\rm scalar},
\ee 
where the limit taken on the left is on the vector rate for $p = 4$ while the rates on the right side are for $p = 2$.   One can check that this holds indeed, a non-trivial test! Here we have used $|\beta| =  \nu^{2}_{1}/e^{2} \to 0$ and $\bar\nu_{0} = e \sqrt{\alpha}$.  Note  $\beta = 0$,  from (\ref{alpha-beta-r}), gives ${\rm tr} F^{4} = ({\rm tr} F^{2})^{2}/2 = 2 \alpha^{2}$ with $\alpha = {\rm tr} F^{2}/2$.\\
\noindent
\textbf{The} $\mathbf {p = 5}$ \textbf{or 6 case:}  For either case, the rate is simply given by (\ref{text-field-pp-rate}) and when expressed in Lorentz invariants $\alpha, \beta$ and $\gamma$ given in (\ref{abc}) via (\ref{parameter-relation}), we have the rate 
\bea\label{5-6-rate}
&&{\cal W}^{(\rm String)}_{p, \,p} = \frac{\left[e\sqrt{\alpha}\right]^{\frac{p - 3}{2}} \, e^{2} \sqrt{\beta \gamma }} { (2 \pi)^{p - 2}}   \,e^{- \frac{ \pi m^{2}}{e \sqrt{\alpha}}}\nn
&\,& \qquad \times \frac{\left[\cosh \pi \sqrt{\frac{|\beta|}{\alpha} } + \cosh \pi \sqrt{\frac{|\gamma|}{\alpha}} \right]^{2}}{\sinh \pi \sqrt{\frac{|\beta|}{\alpha}} \sinh\pi\sqrt{\frac{|\gamma|}{\alpha}}}.
 \eea
For $p = 5$, we have the explicit Lorentz invariant rate
\bea\label{5-rate}
&&{\cal W}^{(\rm String)}_{5, 5} = \nn
&& \frac{e^{3 }\sqrt{\alpha\beta\gamma}} { (2 \pi)^{3} }\, 
  \frac{\left[\cosh \pi \sqrt{\frac{|\beta|}{\alpha}} + \cosh \pi \sqrt{\frac{|\gamma|}{\alpha}} \right]^{2}}{\sinh\pi \sqrt{\frac{|\beta|}{\alpha}}\, \sinh\pi\sqrt{\frac{|\gamma|}{\alpha}}} \,e^{- \frac{ \pi m^{2}}{e \sqrt{\alpha} }}.~\quad
\eea
By setting $F_{ab} = 0$ (giving $\beta = 0$ and $\gamma = 0$) with $a, b = 1, \cdots 5$, we have, from (\ref{5-rate}) and (\ref{dof-c}),  
\bea\label{5-QED}
{\cal W}^{(\rm String)}_{5, 5}  &=&  16 \frac{(e E)^{3}}{(2 \pi)^{5}} \, e^{- \frac{ \pi m^{2}}{e E }} = 16\, {\rm W}^{\rm (QED)}_{\rm scalar},\nn
{\cal W}^{\rm (QED)}_{\rm spinor} &=& 4 {\cal W}^{\rm (QED)}_{\rm scalar}, \, {\cal W}^{\rm (QED)}_{\rm vector} = 5 {\cal W}^{\rm (QED)}_{\rm scalar},~~~\quad 
\eea
where  $\alpha = E^{2}$ with $E = |\vec E|$ and $E_{a} =  F_{0 a}$.  Further, from Table \ref{table1}, we have in general
\be\label{5-string-QED}
{\cal W}^{(\rm String)}_{5, 5} = 3 {\cal W}^{\rm (QED)}_{\rm scalar}  + 2 {\cal W}^{\rm (QED)}_{\rm spinor} + {\cal W}^{\rm (QED)}_{\rm vec.}.
\ee
We re-express the rate  (\ref{5-rate}) as
\bea\label{5-string}
&&{\cal W}^{(\rm String)}_{5, 5} = \frac{e^{3 }\sqrt{\alpha\beta\gamma}} { 4 (2 \pi)^{3} } \,{\rm csch}  \pi \sqrt{\frac{|\beta|}{\alpha}} {\rm csch} \pi\sqrt{\frac{|\gamma|}{\alpha}}\nn
&&\times \,e^{- \frac{ \pi m^{2}}{e \sqrt{\alpha} }}\,\left\{ 3 + 8 \cosh \pi \sqrt{\frac{|\beta|}{\alpha}} \cosh \pi \sqrt{\frac{|\gamma|}{\alpha}}  \right.\nn
&&\left. + 1 + 2 \left[\cosh 2 \pi \sqrt{\frac{|\beta|}{\alpha}}  + \cosh 2 \pi \sqrt{\frac{|\gamma|}{\alpha}}\right]\right\}.
\eea
Combining  (\ref{5-string-QED}), (\ref{5-QED}) and (\ref{5-string}),  we have now the explicit Lorentz invariant QED rates for a pair of scalars, a pair of spinors and a pair of vectors, respectively, as
\bea\label{5-rates}
&&{\cal W}^{\rm (QED)}_{\rm scalar} = \frac{e^{3 }\sqrt{\alpha\beta\gamma}} { 4 (2 \pi)^{3}} {\rm csch} \pi \sqrt{\frac{|\beta|}{\alpha}} \,{\rm csch}\pi\sqrt{\frac{|\gamma|}{\alpha}}  e^{- \frac{ \pi m^{2}_{\rm scalar}}{e \sqrt{\alpha} }},\nn
&& {\cal W}^{\rm (QED)}_{\rm spinor} = \frac{e^{3 }\sqrt{\alpha\beta\gamma}} { (2 \pi)^{3} }  \coth \pi \sqrt{\frac{|\beta|}{\alpha}}\, \coth \pi\sqrt{\frac{|\gamma|}{\alpha}} \nn
&&~~~~~~~~~~~~~~~\times e^{- \frac{ \pi m^{2}_{\rm spinor}}{e \sqrt{\alpha} }},\nn
&&{\cal W}^{\rm (QED)}_{\rm vec.} = \frac{e^{3 }\sqrt{\alpha\beta\gamma}} { 4 (2 \pi)^{3} }  \, e^{- \frac{ \pi m^{2}_{\rm vec.}}{e \sqrt{\alpha} }}\nn
&&~~~~~~~~~~~~~~\times \,   \frac{1 + 2 \left[ \cosh 2 \pi \sqrt{\frac{|\beta|}{\alpha}}  + \cosh 2 \pi \sqrt{\frac{|\gamma|}{\alpha}}\right] }{\sinh\pi \sqrt{\frac{|\beta|}{\alpha}}\, \sinh\pi\sqrt{\frac{|\gamma|}{\alpha}}}.~~
\eea
For $p = 6$, we have, from (\ref{5-6-rate}), the rate as
\bea \label{6-rate}
{\cal W}^{(\rm String)}_{6, \,6} &=& \frac{\left[e\sqrt{\alpha}\right]^{\frac{3}{2}} \, e^{2} \sqrt{\beta \gamma }} { (2 \pi)^{4}} \nn
&\times& \frac{\left[\cosh \pi \sqrt{\frac{|\beta|}{\alpha} } + \cosh \pi \sqrt{\frac{|\gamma|}{\alpha}} \right]^{2}}{\sinh \pi \sqrt{\frac{|\beta|}{\alpha}} \sinh\pi\sqrt{\frac{|\gamma|}{\alpha}}}  \,e^{- \frac{ \pi m^{2}}{e \sqrt{\alpha}}}.~~~~~~
 \eea
From Table \ref{table1}, we expect to have in general
 \be\label{6-string-QED}
 {\cal W}^{(\rm String)}_{6, \,6} = 2  {\cal W}^{\rm (QED)}_{\rm scalar}  + 2 {\cal W}^{\rm (QED)}_{\rm spinor} + {\cal W}^{\rm (QED)}_{\rm vec.},
 \ee
 and the following, from (\ref{6-rate}), (\ref{dof-c}) and by setting $F_{ab} = 0$ (giving $\beta = \gamma  = 0$) with $a, b = 1, \cdots 6$, 
   \bea\label{6-QED}
  {\cal W}^{(\rm String)}_{6, \,6} &=&  16 \, \frac{\left(e E\right)^{\frac{7}{2}}} { (2 \pi)^{6}} \,e^{- \frac{ \pi m^{2}}{e E}} = 16\, {\cal W}^{\rm (QED)}_{\rm scalar},\nn
  {\cal W}^{\rm (QED)}_{\rm spinor} &=& 4 {\cal W}^{\rm (QED)}_{\rm scalar}, \, {\cal W}^{\rm (QED)}_{\rm vec.} = 6 {\cal W}^{\rm (QED)}_{\rm scalar},\quad
\eea 
where  $E = |\vec E|$ with $E_{a} = F_{0a}$.  We re-write (\ref{6-rate}) as
\bea
&&{\cal W}^{(\rm String)}_{6, 6} = \frac{\left(e \sqrt{\alpha}\right)^{\frac{3}{2}} \, e^2  \sqrt{\beta\gamma}} { 4 (2 \pi)^{4} }  {\rm csch} \pi \sqrt{\frac{|\beta|}{\alpha}} {\rm csch}  \pi\sqrt{\frac{|\gamma|}{\alpha}}\nn
&& \times e^{- \frac{ \pi m^{2}}{e \sqrt{\alpha} }} \, \left\{2 + 8 \cosh \pi \sqrt{\frac{|\beta|}{\alpha}} \cosh \pi \sqrt{\frac{|\gamma|}{\alpha}} \right.\nn
&& \left. + 2 \left[1 + \cosh 2 \pi \sqrt{\frac{|\beta|}{\alpha}}  + \cosh 2 \pi \sqrt{\frac{|\gamma|}{\alpha}}\right]  \right\}.~~~~
\eea
and therefore have the respective Lorentz invariant QED rates as
\bea\label{6-rates}
&&{\cal W}^{\rm (QED)}_{\rm scalar} = \frac{\left[e \sqrt{\alpha}\right]^{\frac{3}{2}} \, e^{2} \sqrt{\beta\gamma}} {4 (2 \pi)^{4}}\,{\rm csch} \pi \sqrt{\frac{|\beta|}{\alpha}}\, {\rm csch}  \pi\sqrt{\frac{|\gamma|}{\alpha}} \nn
&&~~~~~~~~~~~~~\times e^{- \frac{ \pi m^{2}_{\rm scalar}}{e \sqrt{\alpha} }},\nn
 && {\cal W}^{\rm (QED)}_{\rm spinor} = \frac{ \left[e \sqrt{\alpha}\right]^{\frac{3}{2}} \, e^{2} \sqrt{\beta\gamma}} {(2 \pi)^{4}}\,  \coth \pi \sqrt{\frac{|\beta|}{\alpha}}\nn
 &&~~~~~~~~~~~~~\times  \coth \pi\sqrt{\frac{|\gamma|}{\alpha}} \,e^{- \frac{ \pi m^{2}_{\rm spinor}}{e \sqrt{\alpha} }},\nn
&&{\cal W}^{\rm (QED)}_{\rm vec.} = \frac{\left(e \sqrt{\alpha}\right)^{\frac{3}{2}} \, e^{2} \sqrt{\beta\gamma}} {2 (2 \pi)^{4}}\,e^{- \frac{ \pi m^{2}_{\rm vec.}}{e \sqrt{\alpha} }}\nn
&& ~~~~~~~~~~~~~\times \frac{1 + \cosh 2 \pi \sqrt{\frac{|\beta|}{\alpha}}  + \cosh 2 \pi \sqrt{\frac{|\gamma|}{\alpha}}}{\sinh\pi \sqrt{\frac{|\beta|}{\alpha}}\, \sinh\pi\sqrt{\frac{|\gamma|}{\alpha}}}.
\eea 
 The above Lorentz invariant QED rates in either $p = 5$ or $p = 6$ meet the non-trivial test, following \cite{note2} and the discussion given right after (\ref{4-rates}).  Again as an illustration, we show this for the vector rate relation between the $p = 6$ and the $p = 4$ when we take $\gamma \to 0$.  We take the limit (\ref{text-rate-relation}) for the $p = 6$ vector rate and have
 \bea
&&\lim_{\gamma \to 0} \frac{(2\pi)^{2}}{e \sqrt{\alpha}} {\cal W}^{\rm (QED)}_{\rm vec.} \nn
&\,& = \frac{\left(e \sqrt{\alpha}\right)^{\frac{3}{2}} e \sqrt{|\beta|}}{(2\pi)^{3}} \, e^{- \frac{ \pi m^{2}_{\rm vec.}}{e \sqrt{\alpha} }} \frac{ 2 + \cosh 2 \pi \sqrt{\frac{|\beta|}{\alpha}}}{\sinh\pi \sqrt{\frac{|\beta|}{\alpha}}} \nn
&\,& = 2 {\cal W}^{\rm (QED)}_{\rm scalar}  + {\cal W}^{\rm (QED)}_{\rm vec.},
 \eea 
 where the scalar rate and the vector rate on the right side of the last equality are for the $p = 4$ case with all the masses identified, the expected indeed. \\ 
\noindent
\textbf{Discussion:}  In this Letter, we report the explicit Lorentz invariant QED pair production rate for a pair of charged particles, which can be scalars, spinors or vectors, in diverse dimensions for the most general constant background of electric and magnetic fields allowed. This is obtained via the so-called field theory limit from the corresponding open string one  for a system of  two Dp branes as specified, with the visible Dp carrying the same background, plus the information about the worldvolume field content and certain properties of the rates.  Except for $p = 1$ case and the scalar and spinor rates in $p = 3$, all the other explicit Lorentz invariant rates have not been given before, to the best of our knowledge. For $p > 3$ (or $d > 4$), the rates cannot be obtained from the usual QED one-loop computations due to the underlying field theory being non renormalisable while the stringy computations, with the field theory limit taken, work fine. The $d = 3$ rate with a tunable magnetic field may be useful in experimentally testing the analogue Schwinger effect using lower dimensional condensed-matter systems (see a recent review  \cite{Fedotov:2022ely} and references therein on this) since the usual 4-dimensional one requires an electric field $\sim 10^{18}$ volts/m which is far from reachable in the foreseeable future. The open string pair production, when viewed from the closed string, corresponds to the generation of closed strings. Taking the respective low energy limit will give rise to the generating rate of gravitational waves. We will report this elsewhere.


\medskip
\noindent
\textbf{Acknowledgments}: We acknowledge the support by grants from the NNSF of China with Grant No: 12275264 and 12247103.

\bigskip

\appendix

\noindent
\textbf{Appendix:} Following\cite{Jia:2019hbr}, the open string one-loop annulus amplitude for the two Dp system under consideration is
\bea\label{t-amplitude-annulus-pp}
&&\frac{\Gamma_{p,p}}{V_{1 + p}} = \frac{ 2^2  \sqrt{\det(\mathbb{I} + \hat F_{p})}}{ (8 \pi^2 \alpha')^{\frac{p + 1}{2}}}  \int_0^\infty \frac{d t}{t^{\frac{p - 3}{2}}} \prod_{\alpha =0 }^{2} \frac{\sin \pi \hat \nu_{\alpha}}{\sinh \pi \hat\nu_{\alpha} t}  \nn
&&\times \left[\sum_{\alpha = 0}^{2}\cosh^{2}\pi \hat \nu_{\alpha} t- 2  \prod_{\alpha = 0}^{2} \cosh\pi \hat\nu_{\alpha} t - 1\right]e^{- \frac{y^2 t}{2\pi\alpha'}} \nn
&& \times \prod_{n=1}^{\infty} Z_{n},
 \eea
 where  $\hat\nu_{0} = i \bar{\hat \nu}_{0}$ and
  \begin{widetext}
 \be
 Z_{n}  = \frac{\prod_{a = 1}^{2} \left[\left(1 - 2 |z|^{2n} \cos \pi \bar{\hat \nu}_{0} t \cosh\pi (\hat \nu_{1} + (-)^{a} \hat\nu_{2}) t + |z|^{4n}\right)^{2 } + 4 |z|^{4n} \sin^{2} \pi \bar{\hat \nu}_{0} t \sinh^{2} \pi (\hat \nu_{1} + (-)^{a} \hat\nu_{2}) t \right]}{(1 - |z|^{2n})^{2} [1 - 2 |z|^{2n} \cos 2 \pi \bar{\hat\nu}_{0} t + |z|^{4n}]  \prod_{\alpha = 1}^{2} [1 - 2 |z|^{2n} \cosh 2 \pi \hat nu_{\alpha} t + |z|^{4n}]}
 \ee
 \end{widetext}
and is completely determined by the dimensionless electromagnetic flux $\hat F$ on the visible brane and the $(1 + p)$ eigenvalues of the following matrix 
 \be\label{matrix|w} 
 w_{\alpha}^{~\beta} = \left[\left(\mathbb{I} - \hat F\right)\left(\mathbb{I}+ \hat F\right)^{-1}\right]_{\alpha}^{~~\beta}.
 \ee
 In the above, $|z| = e^{- \pi t}$.  We have proved in \cite{Jia:2019hbr} that for every eigenvalue  $\lambda$ of $w$, its inverse $\lambda^{-1}$  is also an eigenvalue.  So the $(1 + p)$ eigenvalues $\lambda_{\alpha}$ are pairwise. When $p =$ even, this must imply that one of the eigenvalues is unity.    For convenience, we  relabel the eigenvalues pairwise as $\lambda_{\kappa}$ and $\lambda^{-1}_{\kappa}$ with $\kappa = 0, 1\cdots [(p - 1)/2] $ and keep in mind that there is one additional $\lambda = 1$ when $p =$ even.  The amplitude actually depends only on the sum of $\lambda_{\kappa} + \lambda^{-1}_{\kappa}$ for each $\kappa$.   The eigenvalues $\lambda_{\kappa}$ can be determined by the above matrix $w$  via the relations  listed in Table \ref{table2}\cite{Jia:2019hbr}. 
 \begingroup
 \squeezetable
 \begin{table} [!htbp]
  \caption{\label{table2}The equations determining the corresponding eigenvalues for $1\le p \le 6$.}
\begin{center} 
\begin{tabular}{|c|c|}   
\hline
   \diagbox{$p$}{} & Eigenvalue Eq.\\
 \hline
 1 & $\lambda_{0} + \lambda^{-1}_{0} = {\rm tr} w$\\
 \hline
 2 &$\lambda_{0} + \lambda^{-1}_{0} = {\rm tr} w - 1$,  $\lambda = 1$\\
 \hline
 3 & $\sum_{\alpha =0}^{1} (\lambda_{\alpha} + \lambda^{-1}_{\alpha}) = {\rm tr} w,   \sum_{\alpha =0}^{1} (\lambda^{2}_{\alpha} + \lambda^{-2}_{\alpha}) = {\rm tr} w^{2}$\\
\hline
 4 & $\sum_{\alpha =0}^{1} (\lambda_{\alpha} + \lambda^{-1}_{\alpha}) = {\rm tr} w - 1$, \\
 & $ \sum_{\alpha =0}^{1} (\lambda^{2}_{\alpha} + \lambda^{-2}_{\alpha}) = {\rm tr} w^{2} - 1, \, \lambda  = 1$\\
 \hline 
 5& $\sum_{\alpha =0}^{2} (\lambda_{\alpha} + \lambda^{-1}_{\alpha}) = {\rm tr} w,   \sum_{\alpha =0}^{2} (\lambda^{2}_{\alpha} + \lambda^{-2}_{\alpha}) = {\rm tr} w^{2}$ \\
 & $\sum_{\alpha =0}^{2} (\lambda^{3}_{\alpha} + \lambda^{-3}_{\alpha}) = {\rm tr} w^{3}$\\
 \hline
 6 & $\sum_{\alpha =0}^{2} (\lambda_{\alpha} + \lambda^{-1}_{\alpha}) = {\rm tr} w - 1$, \\
    &$\sum_{\alpha =0}^{2} (\lambda^{2}_{\alpha} + \lambda^{-2}_{\alpha}) = {\rm tr} w^{2} - 1$, \\
   & $\sum_{\alpha =0}^{2} (\lambda^{3}_{\alpha} + \lambda^{-3}_{\alpha}) = {\rm tr}  w^{3} - 1$, \, $ \lambda  = 1$\\
 \hline
 \end{tabular}
\end{center}
 \end{table}
 \endgroup
 We can express each eigenvalue as $\lambda_{\kappa} = e^{2 i \pi \hat{\nu}_{\kappa}}$ and so $ \lambda_{\kappa}  + \lambda^{-1}_{\kappa} = 2 \cos 2\pi {\hat \nu}_{\kappa}$.  Here ${\hat \nu}_{\kappa}$ takes either a real or an imaginary value.  For the former, ${\hat \nu}_{\kappa} \in [0, 1)$.  For the latter,  we have shown in \textbf{Appendix A} of \cite{Jia:2019hbr} that there is at most one imaginary ${\hat \nu}_{\kappa}$.  This occurs only if there is a non-vanishing (in a Lorentz invariant sense) electric field in an otherwise general constant electromagnetic flux.  This is the case we consider in this Letter as already indicated in the amplitude (\ref{t-amplitude-annulus-pp}).  Without loss of generality, we  choose this imaginary eigenvalue as $\hat{\nu}_{0} = i \bar {\hat \nu}_{0}$ with $\bar{\hat \nu}_{0} \in [0, \infty)$.   Note  $\hat F_{\alpha\beta} = 2\pi \alpha' e F_{\alpha\beta}$ with $e$ the usual dimensionless charge unit in natural units, $F_{\alpha\beta}$ the usual electromagnetic tensor and $\alpha, \beta = 0, 1, \cdots, p$. 

Note that the factor $\sin\pi \bar{\hat \nu}_{0} t$ in the denominator of the integrand of the amplitude (\ref{t-amplitude-annulus-pp}) vanishes along the positive $t$-axis at $\pi \bar{\hat \nu}_{0} t_{k} = k \pi$ with $k = 1, 2, \cdots$, giving rise to an infinite number of simple poles of this integrand.  This indicates that the amplitude has an imaginary part and the system decays via the open string pair production to release the excess energy introduced by the electromagetic field on the visible brane.  The decay rate can be computed via ${\cal W}_{p, p}  \equiv -  2 \,{\rm Im} \Gamma_{p, p} / V_{1 + p}$, following \cite{Bachas:1992bh, Porrati:1993qd, Lu:2023jxe, Hattori:2023egw}, and is given 
 \begin{widetext}
 \be\label{decay-rate}
{\cal W}_{p, p} = - \frac{\sqrt{\det(\mathbb{I}+ \hat F_{p})} \sinh \pi \bar{\hat \nu}_{0} \prod_{a = 1}^{2}\sin\pi\hat\nu_{a}}{ 2^{-3}\bar {\hat \nu}_{0} (8 \pi^2 \alpha')^{\frac{p + 1}{2}}} \sum_{k = 1}^{\infty} (-)^{k} \left[\frac{\bar{\hat \nu}_{0}}{k}\right]^{\frac{p - 3}{2}} e^{- \frac{k y^2 }{2\pi\bar{\hat \nu}_{0} \alpha'}} \frac{\left[\cosh\pi \frac{k \hat\nu_{1}}{\bar{\hat \nu}_{0}} - (-)^{k} \cosh\pi \frac{k \hat \nu_{2}}{\bar{\hat \nu}_{0}} \right]^{2}}{\sinh\frac{k \pi\hat \nu_{1}}{\bar{\hat \nu}_{0}} \sinh\frac{k\pi\hat \nu_{2}}{\bar{\hat \nu}_{0}}}  Z_{k} \quad
\ee
where 
\be\label{Znk}
Z_{k} = \prod_{n = 1}^{\infty}\frac{[1 - 2 (-)^{k} |z_{k}|^{2n} \cosh\frac{k \pi(\nu_{1} + \nu_{2})}{\bar\nu_{0}} + |z_{k}|^{4n}]^{2}
[1 - 2 (-)^{k}|z_{k}|^{2n} \cosh\frac{k\pi(\nu_{1} - \nu_{2})}{\bar\nu_{0}} + |z_{k}|^{4n}]^{2} }{(1 - |z_{k}|^{2n})^{4} 
[1 - 2 |z_{k}|^{2n} \cosh \frac{2k\pi\nu_{1}}{\bar\nu_{0}}  + |z_{k}|^{4n}] [1 - 2 |z_{k}|^{2n} \cosh \frac{2k\pi\nu_{2}}{\bar\nu_{0}}  + |z_{k}|^{4n}]},
 \ee  
  \end{widetext} 
 with $|z_{k}| = e^{- k \pi/\bar{\hat\nu}_{0}}$.
 The pair production rate corresponds to the first $k = 1$ term of the above decay rate, following \cite{nikishov, Cohen:2008wz} (see also the recent discussion \cite{Lu:2023jxe} on this ), and is given for $1 \le p \le 6$ as
 \bea\label{pp-rate}
&&{\cal W}^{(\rm String)}_{p,\, p} = \frac{ 2^3 \, \sqrt{\det(\mathbb{I} + \hat F_{p}) } \sinh \pi \bar{\hat \nu}_{0} \sin\pi{\hat \nu}_{1} \sin\pi{\hat \nu}_{2}}{\bar {\hat \nu}_{0} (8 \pi^2 \alpha')^{\frac{p + 1}{2}}}  \nn
&&\qquad \times \bar{\hat \nu}_{0}^{\frac{p - 3}{2}} e^{- \frac{ y^2 }{2\pi\bar\nu_{0} \alpha'}} \frac{\left[\cosh \frac{\pi {\hat \nu}_{1}}{\bar{\hat \nu}_{0}} + \cosh \frac{ \pi {\hat \nu}_{2}}{\bar{\hat \nu}_{0}} \right]^{2}}{\sinh\frac{ \pi{\hat \nu}_{1}}{\bar{\hat \nu}_{0}} \sinh\frac{\pi{\hat \nu}_{2}}{\bar{\hat \nu}_{0}}}  Z_{1},
 \eea
 where $Z_{1}$ is given by (\ref{Znk}) for $k = 1$.  
 
 We take $p = 5 \,\,{\rm or} \,\, 6$ as an illustration on how to determine the $\hat{\nu}_{\kappa}$ from the eigenvalue equations given in Table \ref{table2}. The lower-$p$ cases can be obtained from either of these two by taking the respective limits given later on.  Now from (\ref{matrix|w}), we have $w = 2 (\mathbb{I} + \hat F)^{-1} - \mathbb{I},\, w^{2} = 4 (\mathbb{I} + \hat F)^{-2}  - 4 (\mathbb{I} + \hat F)^{-1} + \mathbb{I}$ and $w^{3} = 8 (\mathbb{I} + \hat F)^{-3} - 12 (\mathbb{I} + \hat F)^{- 2} + 6 (\mathbb{I} + \hat F)^{-1}  - \mathbb{I}$.
 Note that for either case we have $\det (\mathbb{I} + \hat F) =   (1 - \hat \alpha) (1 - \hat\beta) (1 - \hat\gamma)$ with 
 \bea\label{hat-abc-r}
 &&\hat \alpha + \hat\beta + \hat \gamma = \frac{1}{2} {\rm tr} \hat F^{2},\nn
 && \hat \alpha (\hat \beta + \hat \gamma) + \hat\beta \hat\gamma = - \frac{1}{4} \left[{\rm tr} \hat F^{4} - \frac{1}{2} ({\rm tr} \hat F^{2})^{2}\right], \nn
 && \hat \alpha \hat\beta\hat\gamma = \frac{1}{3!} \left[{\rm tr} \hat F^{6} - \frac{3}{4} {\rm tr} \hat F^{4} {\rm tr} \hat F^{2} + \frac{1}{8}({\rm tr} \hat F^{2})^{3}\right].\quad
 \eea
From these, one can solve to give ${\rm tr} \hat F^{4} = 2 (\hat\alpha^{2} + \hat\beta^{2} + \hat\gamma^{2})$ and ${\rm tr} \hat F^{6} = 2 (\hat\alpha^{3} + \hat\beta^{3} + \hat\gamma^{3})$. In the above ${\rm tr} {\hat F}^{2n} \equiv ({\hat F}^{2n})_{\alpha}^{\,\,\alpha}$ and ${\rm tr} \hat F^{2n + 1} = 0$ due to $\hat F$ being antisymmetric.   One can show for either case $({\hat F}^{6})_{\alpha}^{\,\,\beta} = \hat\alpha\hat\beta\hat \gamma\, \delta_{\alpha}^{\,\,\beta} - (\hat\alpha\hat\beta + \hat\alpha\hat\gamma + \hat\beta\hat\gamma) (\hat F^{2})_{\alpha}^{\,\,\beta} + (\hat\alpha + \hat\beta + \hat\gamma) (\hat F^{4})_{\alpha}^{\,\,\beta}$.  Using this and by mathematical induction, one can show in general ${\rm tr} \hat F^{2n} = 2 (\hat\alpha^{n} + \hat \beta^{n} + \hat\gamma^{n})$.  We have then ${\rm tr} (\mathbb{I} + \hat F)^{-1} = \sum_{n = 0}^{\infty} {\rm tr} \hat F^{2n} = 2[(1 - \hat\alpha)^{-1} + (1 - \hat\beta)^{- 1} + (1 - \hat\gamma)^{-1}],  \,{\rm tr} (\mathbb{I} + \hat F)^{-2} = \sum_{n = 0}^{\infty} (2 n + 1) {\rm tr} \hat F^{2n} = 2 [(1 + \hat\alpha)(1 - \hat\alpha)^{-2} + (1 + \hat\beta)(1 - \hat\beta)^{- 2} + (1 + \hat\gamma)(1 - \hat\gamma)^{-2}]$ and ${\rm tr} (\mathbb{I} + \hat F)^{-3} = \sum_{n = 0}^{\infty} (2 n + 1)(n + 1) {\rm tr} \hat F^{2n} = 2 [(1 + 3\,\hat\alpha)(1 - \hat\alpha)^{- 3} + (1 + 3\, \hat\beta)(1 - \hat\beta)^{- 3} + (1 + 3 \,\hat\gamma)(1 - \hat\gamma)^{- 3}]$. 
 
 With the above,  we can compute ${\rm tr} w, {\rm tr} w^{2}$ and ${\rm tr} w^{3}$ in terms of the Lorentz invariants $\hat\alpha, \hat\beta$ and $\hat\gamma$.  The corresponding eigenvalue equations list in Table {\ref{table2} give, in terms of $\bar{\hat \nu}_{0}$ and $\hat\nu_{a}$, as $\cosh^{2}\pi\bar{\hat \nu}_{0} + \sum_{a = 1}^{2} \cos^{2} \pi\hat\nu_{a} = (1 - \hat\alpha)^{-1}+ (1 - \hat\beta)^{-1} + (1 - \hat \gamma)^{-1}, \,\cosh^{4}\pi\bar{\hat \nu}_{0} + \sum_{a = 1}^{2} \cos^{4} \pi\hat\nu_{a} = (1 - \hat\alpha)^{ - 2} + (1 - \hat\beta)^{- 2} + (1 - \hat \gamma)^{- 2}, \,\cosh^{6}\pi\bar{\hat \nu}_{0} + \sum_{a = 1}^{2} \cos^{6} \pi\hat\nu_{a} = (1 - \hat\alpha)^{ - 3} + (1 - \hat\beta)^{- 3} + (1 - \hat \gamma)^{- 3}$.  These equations must imply that one of the $\hat \alpha, \hat\beta, \hat\gamma$ is positive while the rest two are less or equal zero. Without loss of generality, we choose $\hat\alpha > 0$ and $\hat\beta, \hat\gamma \le 0$ and have the solutions
 \bea\label{hat-parameters}
\sinh \pi \bar{\hat \nu}_{0} &=& \sqrt{\frac{\hat \alpha}{1 -\hat\alpha}}, \, \sin \pi \hat\nu_{1} = \sqrt{\frac{ - \hat \beta}{1 - \hat \beta}},\nn
 \sin\pi\hat\nu_{2} &=&  \sqrt{\frac{ - \hat \gamma}{1 - \hat \gamma}}.
\eea  
The rate (\ref{pp-rate}) works for either  $p = 5\,\, {\rm or}\,\, 6$ case when both $\hat{\nu}_{1} \neq 0$ and $\hat{\nu}_{2} \neq 0$. The $p = 3\,\, {\rm or}\, \, 4$ rate can be obtained by taking $\hat{\nu}_{2} \to 0 \,\, ({\rm or}\,\, \hat\gamma \to 0)$ limit while the $p = 1\,\, {\rm or}\,\, 2$ rate can be obtained by taking both $\hat{\nu}_{2} \to 0$ and $\hat{\nu}_{1}  \to 0$ (or both $\hat\gamma$ and $\hat\beta \to 0$) limits, respectively.
 
We now take the so-called field theory limit, $|\hat F_{\alpha\beta}| = 2 \pi \alpha' e |F_{\alpha\beta}| \ll 1$, which gives $\hat \alpha, |\hat\beta|, |\hat\gamma| \ll 1$ from their relations with ${\rm tr} \hat F^{2n}$ ($n = 1, 2, 3$) given earlier.  From (\ref{hat-parameters}), this further implies $\bar{\hat \nu}_{0}, \hat \nu_{a} \ll 1$ with $a = 1, 2$.  We can now have $\pi \bar{\hat \nu}_{0} \approx \sqrt{\hat \alpha} \equiv  2 \pi \alpha' e \sqrt{\alpha} \ll 1, \pi \hat\nu_{1} \approx \sqrt{|\hat \beta|} \equiv 2 \pi \alpha' e \sqrt{|\beta|} \ll 1, \pi \hat \nu_{2} \approx \sqrt{|\hat \gamma|} \equiv 2 \pi \alpha' e \sqrt{|\gamma|} \ll 1$ with $\alpha, \beta, \gamma$ satisfying (\ref{abc}).  In this limit, it is clear $|z_{1}| \to 0$ and $Z_{1} \to 1$, and the resulting rate counts the contribution of the 16 pairs of the lowest open/anti open string modes. If setting $\bar {\hat \nu}_{0} = 2 \alpha' \bar \nu_{0} $ and $\hat \nu_{a} = 2 \alpha' \nu_{a}$, we therefore have from (\ref{pp-rate}) the rate (\ref{text-field-pp-rate}) which is our starting point of this Letter. \\


\begin{thebibliography}{99}

\bibitem{Cheung:2024uhn}
C.~Cheung, A.~Hillman and G.~N.~Remmen, ``Bootstrap Principle for the Spectrum and Scattering of Strings,''
 Phys.  Rev.  Lett. \textbf{133},  no.25,  251601  (2024)
doi:10.1103 /PhysRevLett.133.251601 [arXiv:2406.02665 [hep-th]].

\bibitem{Cheung:2025tbr}
C.~Cheung, G.~N.~Remmen, F.~Sciotti and M.~Tarquini,
``Strings from Almost Nothing,''
Phys. Rev. Lett. \textbf{136}, no.25, 251601 (2026)
doi:10.1103/cw4p-cqh7
[arXiv:2508.09246 [hep-th]].

\bibitem{Schwinger:1951nm}
  J.~S.~Schwinger,
  ``On gauge invariance and vacuum polarization,''
  Phys.\ Rev.\  {\bf 82}, 664 (1951).
  
\bibitem{Lu:2023sag}
J.~X.~Lu, ``The open string pair production, its enhancement and the physics behind,''
Phys. Lett. B \textbf{848}, 138397 (2024) doi:10.1016 /j.physletb.2023.138397
[arXiv:2310.07960 [hep-th]].
  
  \bibitem{nikishov}
  A. I. Nikishov, Sov. Phys. JETP 30, 660 (1970); Nucl. Phys. B21, 346 (1970).
  
\bibitem{Kruglov:2001cx} 
  S.~I.~Kruglov,
  ``Pair production and vacuum polarization of vector particles with electric dipole moments and anomalous magnetic moments,''
  Eur.\ Phys.\ J.\ C {\bf 22}, 89 (2001)
  [hep-ph/0110100].  

\bibitem{note3}
In the absence of electromagnetic fields on the D$p$ branes, the mass spectrum for the open string connecting the two D$p$ is $\alpha' M^{2} = -\alpha' p^{2}$ and given as
 \be\label{mass-level}
~~~\alpha' M^{2}  =  \left\{\begin{array}{cc}
\frac{y^{2}}{4\pi^{2} \alpha'} + N_{\rm R} &  (\rm R-sector),\\
\frac{y^{2}}{4\pi^{2} \alpha'}  + N_{\rm NS}  - \frac{1}{2} & (\rm NS-sector),
\end{array}\right.
\ee
where $p = (k, 0)$ with $k$ the momentum along the brane worldvolume directions, $N_{\rm R}$ and $N_{\rm NS}$ are the standard number operators in the R-sector and NS-sector, respectively, as $N_{\rm R} = \sum_{n = 1}^{\infty} (\alpha_{- n} \cdot \alpha_{n} + n d_{-n} \cdot d_{n})$ and $N_{\rm NS} = \sum_{n = 1}^{\infty} \alpha_{- n} \cdot \alpha_{n} + \sum_{r = 1/2}^{\infty} r d_{- r} \cdot d_{r}$.  The R-sector gives fermions with $N_{\rm R} \ge 0$ while the NS-sector gives bosons with $N_{\rm NS} \ge 1/2$.  The $N_{\rm R} = 0$, $N_{\rm NS} = 1/2$ give the usual
massless $4 (8_{\rm F} + 8_{\rm B})$ degrees of freedom(DOF) when $y = 0$ for the two Dp system. 
%

\bibitem{Bachas:1992bh}
  C.~Bachas and M.~Porrati,
  ``Pair creation of open strings in an electric field,''
  Phys.\ Lett.\  B {\bf 296}, 77 (1992)
  [arXiv:hep-th/9209032].
  
\bibitem{Porrati:1993qd}
M.~Porrati,
``Open strings in constant electric and magnetic fields,''
[arXiv:hep-th/9309114 [hep-th]].

\bibitem{Cohen:2008wz}
T.~D.~Cohen and D.~A.~McGady,
``The Schwinger mechanism revisited,''
Phys. Rev. D \textbf{78}, 036008 (2008)
doi:10.1103/PhysRevD.78.036008
[arXiv:0807.1117 [hep-ph]].
    
\bibitem{Lu:2023jxe}
J.~X.~Lu, ``Understanding the open string pair production of the Dp/D0 system,''
JHEP \textbf{11}, 019 (2023)
doi:10.1007 /JHEP11(2023)019
[arXiv:2307.06594 [hep-th]].
  
\bibitem{Jia:2019hbr}
Q.~Jia, J.~X.~Lu, Z.~Wu and X.~Zhu, ``On D-brane interaction {\textbackslash}{\&} its related properties,''
Nucl. Phys. B \textbf{953}, 114947 (2020) doi:10.1016 /j.nuclphysb.2020.114947
[arXiv:1904.12480 [hep-th]]. 

\bibitem{Gavrilov:1996pz}
S.~P.~Gavrilov and D.~M.~Gitman,
``Vacuum instability in external fields,''
Phys. Rev. D \textbf{53}, 7162-7175 (1996)
doi:10.1103/PhysRevD.53.7162
[arXiv:hep-th/9603152 [hep-th]].


\bibitem{Huet:2010nt}
I.~Huet, D.~G.~C.~McKeon and C.~Schubert,
``Euler-Heisenberg lagrangians and asymptotic analysis in 1+1 QED, part 1: Two-loop,''
JHEP \textbf{12}, 036 (2010)
doi:10.1007/JHEP12(2010)036
[arXiv:1010.5315 [hep-th]].

\bibitem{Gusynin:1998bt}
V.~P.~Gusynin and I.~A.~Shovkovy,
``Derivative expansion of the effective action for QED in (2+1)-dimensions and (3+1)-dimensions,''
J. Math. Phys. \textbf{40}, 5406-5439 (1999)
doi:10.1063/1.533037
[arXiv:hep-th/9804143 [hep-th]].

\bibitem{Allor:2007ei}
D.~Allor, T.~D.~Cohen and D.~A.~McGady,
``The Schwinger mechanism and graphene,''
Phys. Rev. D \textbf{78}, 096009 (2008)
doi:10.1103 /PhysRevD.78.096009
[arXiv:0708.1471 [cond-mat.mes-hall]].

\bibitem{Daugherty:1976mg}
J.~K.~Daugherty and I.~Lerche,
``Theory of Pair Production in Strong Electric and Magnetic Fields and Its Applicability to Pulsars,''
Phys. Rev. D \textbf{14}, 340-355 (1976)
doi:10.1103/PhysRevD.14.340

\bibitem{Kim:2003qp}
S.~P.~Kim and D.~N.~Page,
``Schwinger pair production in electric and magnetic fields,''
Phys. Rev. D \textbf{73}, 065020 (2006)
doi:10.1103/PhysRevD.73.065020
[arXiv:hep-th/0301132 [hep-th]].
  
\bibitem{Korwar:2018euc}
M.~Korwar and A.~M.~Thalapillil,
``Finite temperature Schwinger pair production in coexistent electric and magnetic fields,''
Phys. Rev. D \textbf{98}, no.7, 076016 (2018)
doi:10.1103/PhysRevD.98.076016
[arXiv:1808.01295 [hep-th]].

\bibitem{Dunne:2025cyo}
G.~V.~Dunne,
``Heisenberg-Euler and the quantum dilogarithm,''
Phys. Rev. D \textbf{113}, no.8, 8 (2026)
doi:10.1103/ly3t-zp5l
[arXiv:2512.14915 [hep-th]].

\bibitem{Gupta:2025srv}
D.~Gupta and A.~M.~Thalapillil,
``Resurgence in scalar and spinor QED: the Euler{\textendash}Heisenberg Lagrangian in parallel field backgrounds,''
Eur. Phys. J. C \textbf{86}, no.7, 750 (2026)
doi:10.1140/epjc/s10052-026-16024-0
[arXiv:2512.22775 [hep-th]].

\bibitem{Cho:2000ei}
Y.~M.~Cho and D.~G.~Pak,
``Effective action: A Convergent series of QED,''
Phys. Rev. Lett. \textbf{86}, 1947-1950 (2001)
doi:10.1103/ PhysRevLett.86.1947
[arXiv:hep-th/0006057 [hep-th]].

\bibitem{Hattori:2023egw}
K.~Hattori, K.~Itakura and S.~Ozaki,
``Strong-field physics in QED and QCD: From fundamentals to applications,''
Prog. Part. Nucl. Phys. \textbf{133}, 104068 (2023)
doi:10.1016/j.ppnp.2023.104068
[arXiv:2305.03865 [hep-ph]].

\bibitem{note1}
For $d = 1 + p > 4$, the underlying QED is not renormalisable and the usual one-loop computation approach is not applicable. 
An alternative has been taken in that special sets of (in and out) exact solutions of the Dirac equation are constructed in a certain  frame of reference for the external fields for convenience  \cite{Gavrilov:1996pz} and as such the pair production rates can be computed.  For pure electric field, the spinor and scalar rates computed in this Letter agree with those given in \cite{Gavrilov:1996pz} for $d = 5, 6, 7$ cases.  In the presence of magnetic field, however, due to the special choice of the frame of reference in  \cite{Gavrilov:1996pz}, the rates given there are not in an explicit Lorentz invariant form and cannot be directly compared with ours here.
%

\bibitem{note2}
In addition we expect that the relation (\ref{text-rate-relation}) holds also for various QED rates when the proper consideration is taken. Note that a charged scalar in $(3 + p)$ dimensions corresponds to a charged scalar in $(1 + p)$ dimension while the charged spinor correspondence should follow what is given in Table \ref{table1}.  A charged vector in $(3 + p)$ dimensions  gives a charged vector for $p > 1$ or a charged scalar for $p = 1$ plus two charged scalars in $(1 + p)$ dimensions. One can check that this is indeed true when the respective masses are identified. 
%

  
\bibitem{Fedotov:2022ely}
A.~Fedotov, A.~Ilderton, F.~Karbstein, B.~King, D.~Seipt, H.~Taya and G.~Torgrimsson,
``Advances in QED with intense background fields,''
Phys. Rept. \textbf{1010}, 1-138 (2023)
doi:10.1016/j.physrep.2023.01.003
[arXiv:2203.00019 [hep-ph]].
  
  \end{thebibliography}
\end{document}